\documentclass[sigconf]{acmart}

\AtBeginDocument{%
  \providecommand\BibTeX{{%
    \normalfont B\kern-0.5em{\scshape i\kern-0.25em b}\kern-0.8em\TeX}}}

\copyrightyear{2023} 
\acmYear{2023} 
\setcopyright{rightsretained} 
\acmConference[CHI '23]{Proceedings of the 2023 CHI Conference on Human Factors in Computing Systems}{April 23--28, 2023}{Hamburg, Germany}
\acmBooktitle{Proceedings of the 2023 CHI Conference on Human Factors in Computing Systems (CHI '23), April 23--28, 2023, Hamburg, Germany}\acmDOI{10.1145/3544548.3581268}
\acmISBN{978-1-4503-9421-5/23/04}

\newenvironment{myquote}%
  {\list{}{\leftmargin=0.2in\rightmargin=0in}\item[]}%
  {\endlist}

\usepackage{scalerel}
\usepackage{multirow}
\usepackage{enumitem}
\usepackage{xcolor}
\usepackage{listings}
\usepackage{xparse}
\usepackage[utf8]{inputenc}
\usepackage{xcolor}

\usepackage{hyperref}
\usepackage[nameinlink]{cleveref}

\definecolor{code}{HTML}{6A1B9A}
\newcommand{\feedback}[2][]{\emph{``#2''}}
\newcommand{\system}{\textsc{zeno}}
\newcommand{\python}[1]{{\color{code}\lstinline{#1}}}

\newcommand{\decorator}[1]{\python{@#1}}

\newcommand{\model}{\decorator{model}}
\newcommand{\distill}{\decorator{distill}}
\newcommand{\metric}{\decorator{metric}}
\newcommand{\transform}{\decorator{transform}}

\begin{document}

\title{Zeno: An Interactive Framework for Behavioral Evaluation of Machine Learning}

\author{Ángel Alexander Cabrera}
\orcid{0000-0003-0348-3362}
\affiliation{%
  \institution{Carnegie Mellon University}
  \city{Pittsburgh}
  \state{Pennsylvania}
  \country{USA}
}
\author{Erica Fu}
\orcid{0000-0002-9284-5750}
\affiliation{%
  \institution{Carnegie Mellon University}
  \city{Pittsburgh}
  \state{Pennsylvania}
  \country{USA}
}
\author{Donald Bertucci}
\orcid{0000-0002-2726-4108}
\affiliation{%
  \institution{Carnegie Mellon University}
  \city{Pittsburgh}
  \state{Pennsylvania}
  \country{USA}
}
\author{Kenneth Holstein}
\orcid{0000-0001-6730-922X}
\affiliation{%
  \institution{Carnegie Mellon University}
  \city{Pittsburgh}
  \state{Pennsylvania}
  \country{USA}
}
\author{Ameet Talwalkar}
\orcid{0000-0001-6650-1893}
\affiliation{%
  \institution{Carnegie Mellon University}
  \city{Pittsburgh}
  \state{Pennsylvania}
  \country{USA}
}
\author{Jason I. Hong}
\orcid{0000-0002-9856-9654}
\affiliation{%
  \institution{Carnegie Mellon University}
  \city{Pittsburgh}
  \state{Pennsylvania}
  \country{USA}
}
\author{Adam Perer}
\orcid{0000-0002-8369-3847}
\affiliation{%
  \institution{Carnegie Mellon University}
  \city{Pittsburgh}
  \state{Pennsylvania}
  \country{USA}
}

\renewcommand{\shortauthors}{Cabrera et al.}

\begin{abstract}
Machine learning models with high accuracy on test data can still produce systematic failures, such as harmful biases and safety issues, when deployed in the real world.
To detect and mitigate such failures, practitioners run \textit{behavioral evaluation} of their models, checking model outputs for specific types of inputs.
Behavioral evaluation is important but challenging, requiring that practitioners discover real-world patterns and validate systematic failures. 
We conducted 18 semi-structured interviews with ML practitioners to better understand the challenges of behavioral evaluation and found that it is a collaborative, use-case-first process that is not adequately supported by existing task- and domain-specific tools.
Using these findings, we designed \system{}, a general-purpose framework for visualizing and testing AI systems across diverse use cases.
In four case studies with participants using \system{} on real-world models, we found that practitioners were able to reproduce previous manual analyses and discover new systematic failures.
\end{abstract}




\begin{teaserfigure}
  \includegraphics[width=\textwidth]{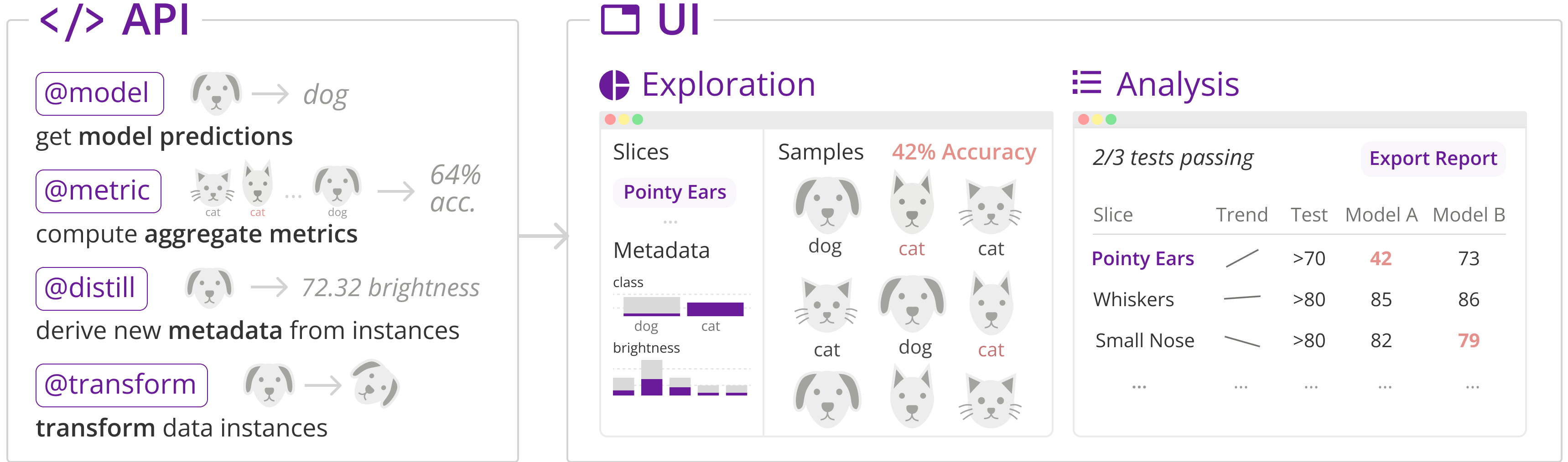}
  \caption{ 
  \system{} is a framework for behavioral evaluation of machine learning (ML) models.
  It has two components, a Python API and an interactive UI.
  The API is used to generate information such as model outputs and metrics.
  Users then interact with the UI to see metrics, create slices, and write unit tests.
  In this toy example, a user is evaluating a cat and dog classifier.
  They see that the model has lower accuracy for dogs with pointy ears, and create a test expecting the slice accuracy to be higher than 70\%.
}
  \Description{Two images of Zeno, the API and the UI. The API shows examples of python decorators that send important information to the UI. The UI shows that a user can create subsets of the data and create tests for performance for future model iterations.}
  \label{fig:teaser}
\end{teaserfigure}

\maketitle

\section{Introduction}
Machine learning (ML) systems deployed in the real world can encode problems such as societal biases \cite{Barocas2016} and safety concerns \cite{NTSB2018}.
Practitioners and researchers continue to discover significant limitations and failures in state-of-the-art models, from systematic misclassification of certain medical images \cite{Oakden-Rayner2020} to racial biases in pedestrian detection models \cite{wilson_predictive_2019}.
In one classic example, \citet{buolamwini_gender_2018} compared the performance of facial classification models across different demographic groups and found that the models performed significantly worse for darker-skinned women compared to lighter-skinned men.

Discovering and validating model limitations is often termed \textit{behavioral evaluation} or testing \cite{Rahwan2019}.
It requires going beyond measuring aggregate metrics, such as accuracy or F1 score, and understanding patterns of model output for subgroups, or slices, of input data.
Enumerating what behaviors a model should have or what types of errors it could produce requires collaboration between stakeholders such as ML engineers, designers, and domain experts \cite{nahar_collaboration_2021, Subramonyam2021Process}.
Behavioral evaluation is also a continuous, iterative process, as practitioners update their models to fix limitations or add features while ensuring that new failures are not introduced \cite{cabrera_what_2022}.

Despite a growing focus on the importance of behavioral evaluation, it remains a challenging task in practice.
Models are often developed without practitioners having clear model requirements or a deep understanding of the products or services in which the model will be deployed \cite{nahar_collaboration_2021}.
Furthermore, many behavioral evaluation tools, such as fairness toolkits, often do not support the types of models, data, and behaviors that practitioners work with in the real world \cite{deng_exploring_2022}.
Practitioners end up manually testing hand-picked examples from users and stakeholders, making it challenging to effectively compare models and pick the best version to deploy \cite{Hopkins2021}.

Given the current state of behavioral evaluation for machine learning, this paper asks two guiding research questions: (1) What are the specific real-world challenges for ML evaluation which are shared across different models, data types, and organizations, and (2) Can an evaluation system addressing these challenges help practitioners discover, evaluate, and track behaviors across diverse ML systems.
To this end, we make the following contributions:

\begin{itemize}[leftmargin=*]
    \item \textbf{Formative study on ML evaluation practices}. 
    Through semi-structured interviews with 18 practitioners, we identify common challenges for behavioral evaluation of ML systems and opportunities for future tools.

    \item  \textbf{\system{}, a general-purpose framework for behavioral evaluation of ML systems}.
    We design and implement a framework for evaluating machine learning models across data types, tasks, and behaviors.
    \system{} (\Cref{fig:teaser}) combines a Python API and interactive UI for creating data slices, exportable reports, and test suites.
    
    \item \textbf{Case studies applying \system{} on diverse models}. 
    We present four case studies of practitioners using \system{} to evaluate their ML systems.
    Using \system{}, practitioners were able to reproduce existing analyses without code, generate hypotheses of model failures, discover and validate new model behaviors, and come up with actionable next steps for fixing model issues.
\end{itemize}

% In summary, we introduce \textbf{\system{}}, a general-purpose framework for behavioral evaluation of machine learning. 
% To design \system{}, we conducted 18 need-finding interviews and evaluated its utility in 3 case studies.
% Lastly, we discuss how \system{} provides a general framework on which future behavioral evaluation tools, such as failure discovery methods, can be built.
% \efcomment{is this useful to have at the end of the paragraph? i feel like if people wanted a summary it would b at the top of the section}
% The practitioners were able to reproduce and discover new model behaviors using \system{} on their own data, and found it to be a powerful and intuitive framework for model evaluation.
% \accomment{maybe a "future direction" sentence like "a common API for discovery methods" or something}

\section{Background and Related Work}

\system{} expands upon work on machine learning evaluation from the fields of human-computer interaction and ML.
We first explore the current state of machine learning evaluation, including common techniques and approaches.
We then describe existing tools for evaluation, and conclude with methods for improving collaboration and shared model understanding in data science and ML.

\subsection{Behavioral Evaluation of Machine Learning}
Evaluating a machine learning model is the challenge of understanding how well a model can accomplish a given task. 
The canonical approach to evaluation is to calculate an aggregate performance metric on a held-out sample of data or test set. 
But just as an IQ test is a rough and imperfect measure of human intellect, aggregate metrics are a rough approximation of model performance. 
They can, for example, hide systematic failures like societal biases, or fail to encode basic capabilities like correct grammar in NLP systems.

To detect and mitigate these important issues, the ML community uses more fine-grained evaluation approaches, often termed \textit{behavioral evaluation} \cite{Rahwan2019, cabrera_what_2022}. 
Inspired by requirements engineering in software engineering, behavioral evaluation focuses on defining and testing the capabilities of an ML system, its expected behavior on a specification of requirements \cite{yang_capabilities_2022, pei_requirements_2022}. 
For example, a practitioner creating a sentiment classification model might check that the model works for double negatives, is invariant to gender, and is accurate for short text. 
In addition to aggregate metrics, they would check how their model performs in these specific scenarios.

A central challenge in behavioral evaluation is deciding \textit{which} capabilities a model should have. 
There can be a practically infinite number of requirements in complex domains, which would be impossible to list and test.
Instead, ML engineers work with domain experts and designers to define the capabilities that a model should have as they iterate on and deploy their ML systems \cite{Subramonyam2021Process}.
As end-users interact with the model in products and services, they also provide feedback on the limitations or expected behaviors that are then used to update the model \cite{Cabrera2021Deblinder}. 

In this work we further explore evaluation in practice through our formative study.
We identify common challenges across domains and opportunities for future tools which we apply when designing and building the \system{} system.

% \system{} is an instantiation of the proposed unified concept of capabilities testing described by \cite{yang_capabilities_2022}\efcomment{do we want to add an explanation here?}. Our formative studies explore the concrete challenges for this type of evaluation which we tackle in \system{}.

\subsection{Model Evaluation Approaches}

There are numerous ML evaluation systems for discovering, validating, and tracking model behaviors \cite{cabrera_what_2022, Rahwan2019}.
The tools use techniques such as visualizations and data transformations to discover behaviors like fairness concerns and edge cases.
\system{} complements some of these systems and integrates the approaches of others.

The behavioral evaluation method most related to \system{} is subgroup, or slice-based, analysis, calculating metrics on subsets of a dataset.
An example tool for slice-based analysis is FairVis \cite{Cabrera2019}, a visual analytics system that allows users to compare subsets of data across metrics to discover intersectional biases.
Errudite \cite{Wu2019} is a similar system for NLP models with which users can create and test subgroups using structured queries.
% Data slices can also be used to define formal tests of model behavior.
% \citet{kang_model_2020-1} applied the concept of software engineering assertions to ML with \textit{model assertions}, expectations of model outputs on slices of input data. 
% \system{} supports a similar type of test which we term \textit{behavioral unit tests}, expectations of model metrics on slices of data. 
% While model assertions are fully programmatic, \system{} tests can be created by nontechnical users with the Analysis UI.
Another common method for behavioral evaluation is metamorphic testing \cite{chen_metamorphic_2019}, a concept from software engineering that involves checking the outputs of a black-box system for inputs that are perturbed in a specific way.
Checklist \cite{Ribeiro2020} is a metamorphic testing tool for NLP models that perturbs text inputs, for example, switching proper nouns and testing if a model's output switches.
% \efcomment{For example, given a sentiment classification model, Checklist will test whether the sentiment of a sentence changes when a proper noun is changed.}
% Another metamorphic testing tool for ML is DeepRoad, which tests self-driving car models by altering images with weather artifacts such as snow \cite{Zhang2018}.
\system{} enables users to do slice-based and metamorphic testing for any domain and task.

A central challenge for behavioral evaluation is \textit{discovering} which behaviors a model has and are important for real-world performance.
Various methods using algorithmic or crowdsourced techniques have shown promise in surfacing such behaviors.
Algorithmic methods are a common approach for detecting groups of instances with high error, often termed ``blindspots''.
SliceFinder is one method that uses metadata to find slices with significantly high loss \cite{chung_slice_2019}.
Often, there is not enough metadata to define slices with high error, so another family of methods uses model embeddings and clustering to find groups with high error \cite{eyuboglu_domino_2022, deon_spotlight_2021}.
Lastly, there are approaches that use end-user reports or crowd feedback to discover model failures or interesting behaviors \cite{Attenberg2011, Cabrera2021Deblinder, Nushi2018}.
% These methods can provide initial subgroups of instances with significant failures, but often generate false positives and unrepresentative hypotheses of behaviors \cite{plumb_evaluating_2022}.
\system{} complements discovery methods by allowing users to formalize, validate, and track hypotheses of systemic errors over time.
% Users can encode insights of potential failures from the above methods and formally validate how present they are for their models using \system{}.
% Furthermore, \system{} indirectly supports discovery by encouraging reuse and sharing of functions.
% For example, if one user creates a \python{@distill} function to measure the brightness of an image, they can share it with others and bootstrap their evaluation process.
% In summary, \system{} complements discovery methods by providing a formal way to encode and validate their hypotheses of model behavior.

Lastly, there are integrated platforms for model evaluation that combine multiple types of analyses.
For instance, Robustness Gym \cite{Goel2021} is a framework for NLP models that supports multiple types of evaluation, including adversarial attacks and robustness checks.
The What-If tool \cite{Wexler2019} is another interactive framework that focuses on using counterfactuals to understand model behavior and fix fairness concerns.
We took a similar approach to these frameworks when designing \system{}, but focused on the more general task of behavioral evaluation for any model or data type.

% While existing systems focus on specific data types, models, or behaviors, \system{} is a framework that generalizes across domains.
% \accomment{is there a stronger claim we can make here that isn't just "we are general"?}\efcomment{also focus on separating the technical(API) and nontechnical(UI) part so that nontechnical people can perform evaluations? Why are these not used in practice? Why should Zeno be used in practice?}.
% By introducing a general paradigm, practitioners only have to learn one tool, can reuse analysis code across models, and can store their analyses in a centralized location for sharing and collaboration.

% \apcomment{Is it useful to allude here why these tools are not used much in practice?  Or at least forward-reference this in interviews... as I think reviewers will want us to differentiate Zeno from these tools}

\subsection{Collaboration and Reporting}

Most ML models are developed by cross-functional teams with stakeholders in technical and non-technical roles. 
While collaboration is essential for deciding how a model should behave and identifying potential failures, there is often limited communication between stakeholders \cite{nahar_collaboration_2021}.
This can lead to unrealistic expectations of model performance or results that do not match designers' expectations.
Multiple methods have been proposed to improve organizations' shared understanding of model behavior.

Interactive systems have shown promise for bridging model knowledge between engineering and other roles.
One example framework, Symphony \cite{bauerle_symphony_2022}, introduces modular data and model analysis components that can be used in both computational notebooks and standalone dashboards to enable more stakeholders to explore model behavior.
Marcelle \cite{francoise_marcelle_2021} similarly uses modular components that allow users to modify an ML pipeline without writing code.

Complex models also require robust reporting methods to ensure that information about data and models is recorded and preserved.
Datasheets for Datasets \cite{gebru_datasheets_2021}, FactSheets \cite{arnold_factsheets_2019}, Nutritional Labels \cite{stoyanovich_nutritional_2019}, and Model Cards \cite{Mitchell2019} codified the first principles for documenting ML details for future use and reproducibility.
Extensions to these reporting methods, namely Interactive Model Cards \cite{crisan_interactive_2022}, have aimed to improve their usability by making them more expressive and interactive.
\system{} is primarily an interactive UI to enable diverse stakeholders to perform model analysis and export results that can be included in reporting methods like model cards.

% \ameet{I'm surprised that you're not mentioning any work on blindspot detection here (though I see you include it later on as future work. It could be worth mentioning here that while there is growing research subcommunity aiming to develop 'robust' ML systems, automated robustness approaches won't replace behavioral testing.  Note that this is often how blindspot detection work is motivated, esp within the ML community, since robust ML is one of the hottest areas within ML right now and so its the first thing that comes to mind for ML folks when hearing about this kind of work}

\section{Formative Interviews with machine learning practitioners}

\begin{table}[b]
    \centering
    \caption{The practitioners in the semi-structured interviews. }
    \begin{tabular}{lll}
        ID & Role & Area \\
        \midrule
        P1 & AI Software Engineer & AI Consulting \\
        P2 & Data Scientist & Clothing Retail \\
        P3 & CTO & Speech Training \\
        P4 & CTO & Voice Assistant \\
        P5 & Senior ML Engineer & Chatbot \\
        P6 & Data Scientist & AI Non-profit \\
        P7 & Data Scientist & Finance \\
        P8 & MS Student & Educational Technology \\
        P9 & ML Engineer & Chatbot \\
        P10 & VP of Data Science & Business Intelligence \\
        P11 & ML Engineer & AI Explainability \\
        P12 & Data Scientist, ML & Ridesharing \\
        P13 & Data Engineer & Educational Technology \\
        P14 & CTO & Health Technology \\
        P15 & CEO & Sensing \\
        P16 & Data Scientist & Search and Recommendation \\
        P17 & ML Research Scientist &  Epidemiology \\
        P18 & Data Scientist & Video Streaming
    \end{tabular}
    \label{tab:studies}
\end{table}

We conducted semi-structured interviews with machine learning practitioners to explore our first research question: What are the common challenges for ML evaluation in practice?
In particular, we aimed to understand the specific challenges practitioners face and the tools they use when evaluating ML models.
The 18 participants, listed in \Cref{tab:studies}, hold various roles related to machine learning development and deployment, from data scientists to CTOs and CEOs of small companies.
The initial participants were recruited through posts on social media networks, e.g., Reddit, LinkedIn, and Discord, and through direct contacts at technology companies. 
Additional participants were then recruited through snowball sampling.
Each interview lasted an hour via video call and participants were compensated with \$20.
The study was approved by our Institutional Review Board (IRB).
% Participants are referenced by their ID (P1 - P18) in the following sections.

Two researchers analyzed the interviews using inductive iterative thematic analysis and affinity diagramming.
From the first few interviews, the researchers extracted common themes around model evaluation, debugging, and iteration, grouping similar findings in an affinity diagram.
After each subsequent interview, the researchers iterated on and refined the themes as needed.
Recruiting for new participants was stopped when no new themes were produced from the last few interviews. 

% \jason{For 3.1, 3.2, 3.3, is it possible to rephrase the subheaders to be the key insight and/or challenge?}

\subsection{Aggregate Metrics Do Not Reflect Model Performance in Deployment} 
% \efcomment{Model performance during development does not reflect performance after deployment}
% OUTLINE
% P1
    % Training/research of models focused on metrics and loss values
    % Often don't translate to real-world performance
    % Track performance on real-world use cases
% P2
    % Often times from end-user reports, dogfooding
    % Run bespoke analyses/tests to verify
    % Put them into their own "test sets" "benchmarks" etc. to track
% P3
    % Existing tools don't support tracking/checking for these issues
    % Specific datatypes/tasks
    % Not compatible/too much effort

All practitioners (18/18) focus on improving aggregate metrics when developing new ML models, but, as P9 admitted, you \feedback[P9]{can perform very well on a training dataset, but when you go to ship the product, it doesn't work nearly as well.}
To ensure that models perform as expected when they are deployed, all practitioners also evaluate their models on real-world use cases.
For example, P16 evaluates their text analysis model on a per-client basis since they had found that their model underperformed for certain types of data, e.g. healthcare notes, that it was not trained on.
This type of behavioral analysis is often also called \textit{qualitative} analysis, looking at specific instances and model outputs to confirm hypotheses of model behavior.

There are various methods practitioners described for discovering model limitations and failures, from end-user reports (see \Cref{sec:collab}) to automated clustering algorithms.
A common technique 11 of the 18 participants mentioned was creating their own data inputs to probe a model and find potential failures, often called ``dogfooding'' in software development.
For example, when selecting an audio transcription service P3 \feedback[P3]{has some data collected we recorded ourselves, and then we pass it to different services and explore the structure of the output} to decide which service provides the qualitatively ``best'' output for their task.
Two participants are exploring automated error discovery methods such as finding clusters with high error or using foundational models \cite{bommasani_opportunities_2022, ribeiro_adaptive_2022} to generate new instances, but still primarily rely on human-generated feedback.

After generating hypotheses of systemic failures, many practitioners craft test sets to validate how prevalent behaviors are (10/18).
The participants had different terms for these sets of instances, including ``golden test sets'', ``dynamic benchmarks'', ``regression tests'', and ``benchmark integration tests''.
Despite the varied terminology, these tests have the same structure: Expectations for model outputs on different subgroups of instances.
For example, P4 has multiple sets of text inputs with common human typos paired with valid outputs that they check before model releases.

None of the participants who conduct this type of behavioral evaluation use standardized frameworks.
This is primarily because existing behavioral evaluation tools do not work for their data or model types, so they develop their own tools, such as scripts or web interfaces, to monitor model performance.
All the participants who do not perform behavioral analyses (8/18) wish to conduct more detailed testing, for example, P1 wants \feedback[P1]{to do some other testing, but we don't do anything because there's not a really easy to set up system to do that}.
Overall, bigger companies are able to dedicate more time to detailed evaluation and building customized tools that smaller companies cannot afford despite their need for more comprehensive evaluation \cite{Hopkins2021}.

\subsection{Challenges in Tracking Continuous Model and Data Updates} 
% \efcomment{Continuous model, data, and distribution changes are hard to keep track of}

% OUTLINE
% P1
    % Update model
    % Get more real-world use cases, data, failures
% P2
    % Stochastic updates, AI services -> don't know how will change
    % AI services
% P3
    % Constantly changing, have to check regressions, etc.
    % Create ways to track these behaviors over model iterations

All practitioners (18/18) we interviewed update their models as they design better architectures, gather more data, and discover real-world use cases and failures.
Participants described this process with different terms, such as ``rapid prototyping'' or ``agile'' methods in which they quickly act on user feedback and deploy updated models.  
P4 and P13 even started with ``wizard-of-oz'' models with a human emulating an AI or non-ML models to gather data and model requirements before developing more complex models.

Although updating a model can improve the overall performance of an ML system, it can also lead to new failures.
This is especially true for stochastic models, such as deep learning, which cannot be deterministically updated.
As P5 lamented, \feedback[P5]{our test set would become so large that if we had to fail for less than 5 [tests] it became super hard to make progress}.
Model updates are even more complicated for teams that rely on external AI services, as practitioners do not control when or how services are updated \cite{chen_did_2021}.
For example, P3's team had to switch their voice-to-text service from Google to Amazon because Google stopped detecting filler words such as `um' after a model update, which was necessary for their product. 

Due to these frequent updates, it becomes important to compare models across important behaviors.
However, since many model evaluations are run inconsistently and across different tools, the history of past performance is often fragmented or lost, making it difficult to find regressions or new failures.

\subsection{Limited Collaboration in Cross-Functional Teams}\label{sec:collab}
% \efcomment{Cross-functional teams are unable to collaborate on model evaluation}

% OUTLINE
% P1
    % ML models built by teams
    % Include various roles, from customer service to ....
% P2
    % Many insights on model failures from users, customer service
    % Test and translate to ML engineers
% P3 
    % Hard to translate failures to be useful for engineers
    % Hard to manually test themselves

\begin{figure*}[t]
  \centering
  \includegraphics[width=\linewidth]{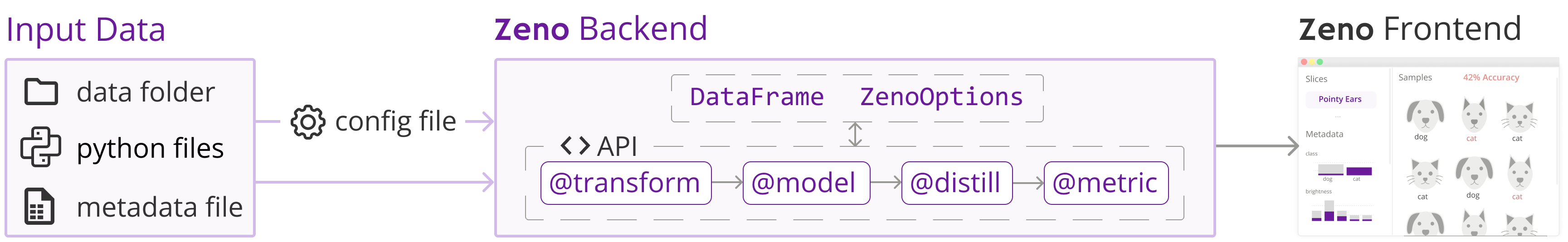}
  \caption{
    \system{}'s architecture overview. 
    The \system{} program and inputs (outlined in {\color{code} purple} boxes) can either be hosted locally or run on a remote machine. 
    \system{} takes a configuration file with information such as paths to data folders, test files, and metadata and creates a parallelized data processing pipeline to run the decorated Python functions.
    The resulting UI is available through an endpoint that can be accessed locally or hosted on a server.
}
  \Description{Diagram of the Zeno Backend and the Zeno Frontend. The diagram shows the Zeno Backend taking input data and the API from the user and generating the Zeno Frontend.}
  \label{fig:api}
\end{figure*}

Modern machine learning development in practice is a collaborative effort that spans different teams and roles.
Each member of a team needs a robust mental model of how an ML system behaves to resolve customer complaints, make management decisions, validate failures, and more.

A common collaboration challenge is making sense of failure reports \cite{Cabrera2021Deblinder} from end-users.
12 of the 18 participants' teams have customer service representatives who parse tickets or complaints from end users and pass them to the engineering teams.
These participants found it challenging to reproduce the reports from end users, which were primarily made up of one-off instances and broad descriptions.
P4's team tackles this challenge with an
\feedback[P4]{internal website where anybody can put potential inputs and expected model outputs} which new models are tested on.

Another collaboration challenge described by 14 participants is communicating model performance with managers and other stakeholders.
For example, P16's management team often makes decisions based solely on a high F1 score, while it is often the case that different clients require different trade-offs between precision and recall.
Many decisions on whether or not to deploy an updated model requires shared knowledge and conversations between engineers, managers, and customers on whether a new model is holistically better than the existing model.

Since engineers often run analyses in ad-hoc scripts or notebooks, knowledge of model behavior can be isolated.
Other stakeholders do not know how a model tends to behave, and can neither make informed decisions on model usage nor provide  information about model errors to engineers for debugging.

\section{Design Goals}

From these interviews and the reviewed studies on ML evaluation, we distilled a set of design goals that a behavioral evaluation system should have. 
The goals focus on general evaluation challenges identified in the formative study, such as defining behaviors and comparing models.
With a system for behavioral evaluation, a user should be able to:

\begin{enumerate}[leftmargin=*]
    \setlength\itemsep{0.5em}
    \item[D1.] \noindent\textbf{Evaluate models with different architectures, tasks, and data types.}
    Machine learning is a broad field with diverse models and tasks ranging from audio transcription to human pose estimation.
    To reduce the learning curve and encourage the reuse of analyses, users should be able to use one framework to perform behavioral evaluations on most ML tasks. 
    
    \item[D2.] \noindent\textbf{Define and measure diverse model behaviors.}
    Model behaviors are varied and complex, from demographic biases to grammatical failures.
    Users should be able to encode most of the behaviors across which they wish to evaluate their models.
    % \jason{maybe work in the terms "flexible" and/or "expressive"? Makes it sound more computer science-y} \accomment{Tried it but it seems a bit system-dependent and I'm trying to keep these user-focused}
    
    \item[D3.] \noindent\textbf{Track model performance over time.}
    Practitioners are continually deploying updated models with new architectures trained on improved data.
    Users should be able to track performance across models and find potential regressions.
    
    \item[D4.] \noindent\textbf{Evaluate model performance without programming.}
    Modern machine learning systems are built by large cross-functional teams with nontechnical users.
    Users should be able to perform behavioral analyses of models without having to write code.

\end{enumerate}

\section{Zeno: An Interactive Evaluation Framework}\label{sec:zeno}

We used these goals to design and implement \system{}, a general-purpose framework for evaluating ML systems across diverse behaviors.
\system{} is made up of two linked components, a Python API and an interactive user interface (UI).
The Python API is used to write functions providing the core building blocks of behavioral evaluation such as model outputs, metrics, metadata, and transformed instances.
Outputs from the API are used to scaffold the interactive UI, which is the primary interface for doing behavioral evaluation and testing.
The \system{} frontend has two primary views: an \textit{Exploration UI} for discovering and creating slices of data and an \textit{Analysis UI} for writing tests, authoring reports, and tracking performance over time \textbf{(D3)}.

Originally, we explored implementing \system{} as either a plugin for computational notebooks or a standalone user interface. 
We decided on a combined programmatic API and interactive UI as we found it could make \system{} both extensible and accessible.
The general Python API allows \system{} to be applied to diverse models, data types, and behaviors \textbf{(D1, D2)}, while the interactive UI allows nontechnical users to run evaluation \textbf{(D4)}.

\system{} is distributed as a Python program.
The Python package includes the compiled frontend which is written in Svelte and uses Vega-Lite \cite{Satyanarayan2017} for visualizations and Arquero~\cite{heer_arquero_2020} for data manipulation.
To run \system{}, users specify settings such as test files, data paths, and column names in a TOML configuration file and launch the processing and UI from the command line (\Cref{fig:api}).
Since \system{} hosts the UI as a URL endpoint, it can either be run locally or run remotely on a server with more compute and still be accessed by users on local machines.
This architecture can scale to large deployed settings and was tested with datasets with millions of instances (e.g. DiffusionDB \cite{wang_diffusiondb_2022}, 2 million images (\Cref{sec:diffusion})).

% \jason{Maybe in this section, can connect to D1 about how it's a separate running system, making it easy to integrate with many kinds of architectures and models. Actually, is that true? Does Zeno have to run on the same machine, b/c of the model? The @model description sort of suggests it does.} \accomment{Jason Wu ran it on a cloud provider and accessed the remote URL to use zeno, if that's what you mean}

% \jason{Around here, say something about lines of code in Zeno} \accomment{what do you mean by lines of code?}
% \jason{How many new lines of code you wrote for Zeno, to give readers a sense of "effort". Note that lines of code is ill-defined, so can just count how many lines of python, html, css, etc separately. For example, "Zeno is distributed as a Python..." it is comprised of about 6000 lines of python code, 4000 lines of html, and 500 lines of css}

\subsubsection{Running example} To explain \system's concepts, we walk through an example use case of a data scientist working at a company deploying a new model.
In the following sections, we use block quotes to show how \system{}'s features would be used in the example.

\begin{myquote}
Emma is a data scientist at a startup developing a voice assistant. 
Her company is using a simple audio transcription model and she has been tasked with understanding how well the model works for their data and what updates they need to make.
\end{myquote}

\subsection{Python API: Extensible Model Analysis}

% \jason{For each of these sections describing key ideas, suggest using a problem - solution format. Have the first sentence / paragraph set up what the problem is, and then the rest describe what the solution is. To avoid being an instruction manual, be sure to convey the rationale for things too, or what alternatives are and why they don't work.}

% OUTLINE
% P1
    % Grand diversity in ML ecosystem
    % Libraries for ML - dozens, different formats, etc. mostly python
    % Data/domains -
    % To generalize as well as possible, Python API
%P2
    % Python API provides functions for extracting core behavioral analysis building blocks
    % Each function same core API - Pandas DataFrame and ZenoOptions object.
% P3
    % First two functions provide core information
    % @predict - model outputs
    % @transform - new inputs
% P4
    % Second two provide analysis info
    % @distill - add metadata columns
    % @metric - calculate metric for subset.
% P4
    % Organized in a data processing pipeline.
    % cached and intelligently run.
    % e.g. parse the distill function to know if dependent on output.

% \jason{I wonder if it's useful to have a system diagram showing (roughly) how people can connect with Zeno. This might also be in your running example too. For example, that it's an independent service that can be run on the same machine or a different machine that builds the model}

A core component of \system{} is an extensible Python API for running model inference and data processing.
The ML landscape is fragmented across many frameworks and libraries, especially for different data and model types.
Despite this fragmentation, most ML libraries are based on Python, so we designed the backend API for \system{} as a set of Python decorator functions that can support most current ML models \textbf{(D1)}.

% We found that we could distill behavioral evaluation to a set of core building blocks for getting model outputs, calculating metrics, and generating new instances and metadata.
% These building blocks can be used to 

\begin{figure}
  \centering
  \includegraphics[width=\linewidth]{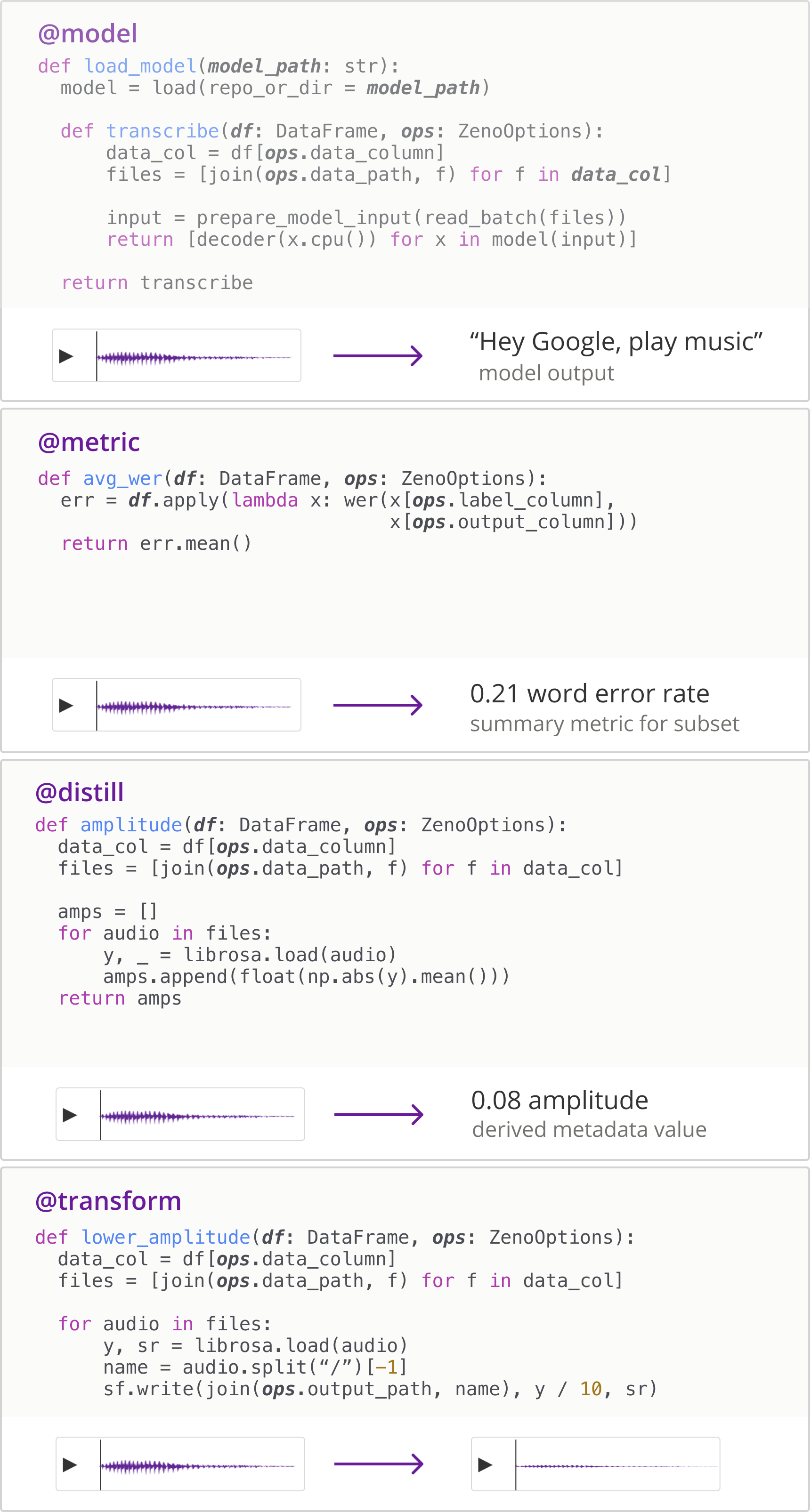}
  \caption{
    The \system{} Python API has four decorator functions: \model{}, \metric{}, \distill{}, and \transform{}. 
    The functions all take the same inputs, a DataFrame and a ZenoOptions object with information such as data paths and column names. 
    \model{} functions return a function for getting running model inference. 
    In the example above, the \model{} function loads a speech-to-text model and returns a function that transcribes audio data.
    \metric{} functions calculate aggregate metrics on subsets of data. 
    Above, the \metric{} function computes the average word error rate (avg\_wer) for transcribed audio. 
    \distill{} functions derive new metadata columns. 
    Above, the \distill{} function calculates the amplitude value from audio. 
    \transform{} functions produce new data inputs. 
    Above, the \transform{} function lowers the amplitude of audio samples. 
    % \system{} then automatically takes the new lowered amplitude data to make text predictions, compute average word error rate, and derive amplitude metadata. 
    %  Users specify predictions (\python{@model}), metrics on data (\python{@metric}), metadata derived from data (\python{@distill}), and transformations on their data (\python{@transform}) through python function. Users can import these function decorators and apply them to their own custom functions. When they apply a decorator, such as \python{@metric}, users get access to parameters like DataFrame and ZenoOptions so they can select data or model outputs. \system{} registers the user decorated functions in the backend and sends the results to the frontend after they are run.
    % \jason{For Fig 3, may want to add comments to the code describing what is going on. Or add to the caption.}\apcomment{Agree this might be helpful.  This code is not very readable to non-Pandas experts (e.g. most of CHI reviewers), so will it scare reviewers into thinking this approach is too hard for anyone to actually use?} \dbcomment{I revised this so the code is not the only thing people rely on.}.
}
  \Description{Four images showing the @model, @metric, @distill, and @transform API with code examples for each. Each image example also includes a real application with speech data for a speech-to-text model.}
  \label{fig:api_example}
\end{figure}

The \system{} Python API (\Cref{fig:api_example}) consists of four decorator functions: \python{@model}, \python{@metric}, \python{@distill}, and \python{@transform}.
We found that these four functions support the building blocks of behavioral evaluation.
All four functions take the same input, a Pandas DataFrame \cite{mckinney_data_2010} with metadata and a \python{ZenoOptions} object.
We chose Pandas as the API for the metadata table due to its popularity, which lowers the learning curve for writing \system{} functions for many data scientists. 
The \python{ZenoOptions} object passes relevant information such as column names and static file paths to the decorated API functions. 
Since \system{} calls API functions dynamically for different models and transformed inputs, \python{ZenoOptions} is necessary for a function to access the correct columns of the DataFrame.

The two core functions that a user must implement to use \system{} are the \python{@model} and \python{@metric} functions.
Functions decorated with \python{@model} return a new function that returns the outputs for a given model.
Since this function is model-agnostic, any ML framework or AI service can be evaluated using \system{} \textbf{(D1)}.
The \python{@metric} decorated functions return a summary number given a subset of data.
\python{@metric} functions can return classic metrics such as accuracy or F1 score, but can also be used for specific tests such as calculating the percentage of changed outputs after data transformations \textbf{(D2)}.

\begin{myquote}
    Emma writes a \python{@model} function which calls her transcription model and returns the transcribed text.
    She then uses a Python library to implement various \python{@metric} functions for common transcription metrics such as word error rate (WER). 
\end{myquote}

\begin{figure*}[t]
  \centering
  \includegraphics[width=\linewidth]{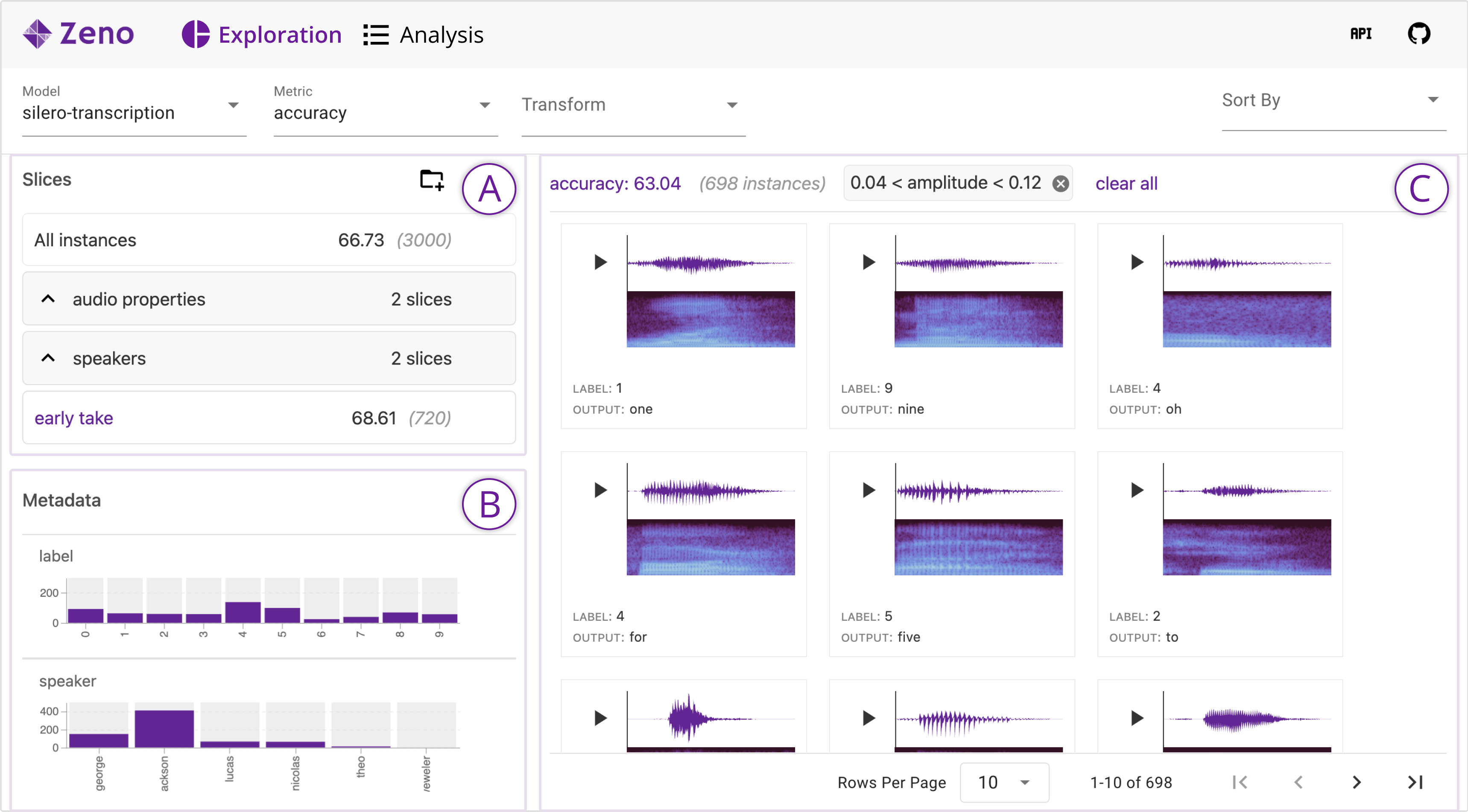}
  \caption{
    The Exploration UI allows users to see data instances and model outputs and investigate model performance.
    In the figure, \system{} is shown for the audio transcription example described in \Cref{sec:zeno}.
    The interface has two components, the Metadata Panel (A \& B) and the Samples View (C).
    The Metadata Panel shows the metadata distributions of the dataset (B) and the slices and folders a user has created (A).
    The metadata widgets are cross-filtered, with the purple bars showing the filtered table distribution.
    The Samples View (C) shows the filtered data instances and outputs, currently those with \textit{0.04 < amplitude < 0.12}, along with the selected metric, in this case, accuracy.
}
  \Description{Image of the Zeno Exploration tab in the UI. This image shows a main data instance view that shows the data itself with its label and prediction. Next to it is a sidebar with histograms for each metadata and created data slices.}
  \label{fig:exploration}
\end{figure*}

\begin{figure*}[t]
  \centering
  \includegraphics[width=\linewidth]{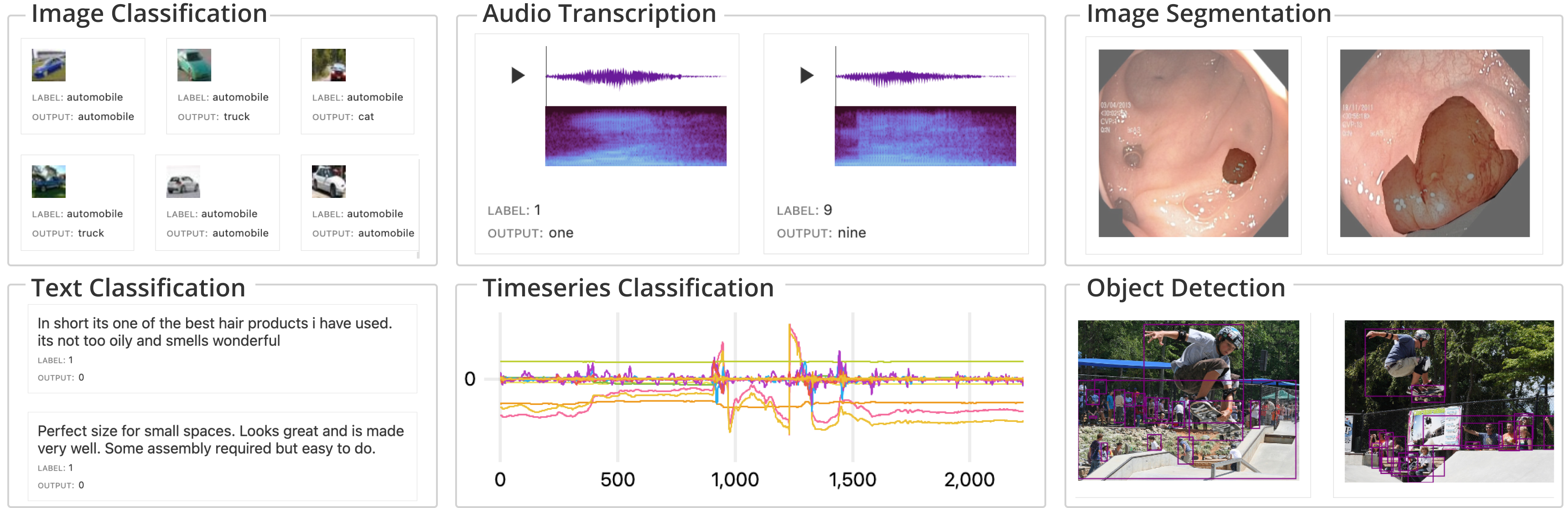}
  \caption{
    The instance view of the Exploration UI (\Cref{fig:exploration}, C) is a modular Python package that can be swapped out for different models and data types.
    New views can be implemented with a single JavaScript file.
    \system{} currently has six implemented views, shown here with the following datasets: image classification (CIFAR-10 \cite{krizhevsky_learning_2009}), audio transcription (Free Spoken Digit Dataset \cite{jackson_jakobovskifree-spoken-digit-dataset_2018}), image segmentation (Kvasir-SEG \cite{ro_kvasir-seg_2020}), text classification (Amazon reviews \cite{ni_justifying_2019}), timeseries classification (MotionSense \cite{malekzadeh_mobile_2019}), and object detection (MS-COCO \cite{fleet_microsoft_2014})
  }
  \Description{Six examples of data views that visualize a specific data instance to be used in Zeno. One for each different type of application: Image Classification, Audio Transcription, Image Segmentation, Text Classification, and Object Detection.}
  \label{fig:views}
\end{figure*}

The two other \system{} decorator functions provide additional functionalities that support behavioral evaluation.
Datasets often do not have sufficient metadata for users to create the specific slices across which they wish to evaluate their models.
For example, a user may want to create a slice for images with low exposure, but most image datasets do not have the exposure level of an image in the metadata.
\python{@distill} decorated functions return a new DataFrame column for a dataset, extracting additional metadata from instances, and allowing users to define more specific slices \textbf{(D2)}.
Users may also want to check the output of their model on modified instances, especially for robustness analyses or metamorphic tests. 
The \python{@transform} function returns a new set of modified instances from a subset of instances. 
For the image exposure example above, a user could write a transformation function that darkens images to check how a model performs for different exposures.
% \efcomment{cases beyond the given dataset? something that shows more the benefit of transformations}.

\begin{myquote}
    Emma knows her users have a range of microphones across which she wants her audio transcription model to work well.
    To test these types of scenarios, she writes a \python{@distill} function that calculates the amplitude of the sound inputs and a \python{@transform} function that adds different types of noise.
\end{myquote}

The \system{} backend builds a data processing pipeline to run the decorated functions and calculate the outputs for the frontend.
For example, \system{} parses the code of each \python{@distill} function to decide whether it depends on model outputs and must be run for each model.
Additionally, \system{} runs the processing and inference functions in parallel, which is especially helpful for transform functions, since each \python{@distill} and \python{@model} function needs to be run on each transformed instance.
Lastly, all \system{} function outputs are cached so any runs after the initial processing are instant.

\subsection{Exploration UI: Create and Track Slices}

% OUTLINE
% P1
    % want to empower non-programmers and other stakeholders to do model analysis.
    % Thus, main interface for Zeno is an interactive UI
    % Python API allows for extensibility, UI for accessibility
% P2
    % Primary view is the exploration UI
    % Two sections, the Slicing Panel and Instance View
% P3
    % Practitioners should directly see the data and model outputs
    % Instance view grid of instances
% P4
    % Behavioral analysis is focused on analyzing subsets of data defined by metadata.
    % Metadata panel shows overviews of metadata
    % Specific views for different metadata types. They crossfilter each other and instance view.
    % Dynamically update metric for subset
% P5
    % Can create slices for interesting subsets
    % Slice creation view WYSIWYG
    % Folders for organization

To empower nontechnical stakeholders to perform behavioral analyses, the main interface of \system{} is an interactive UI  \textbf{(D4)}.
Although the initial \model{} and \metric{} functions are required to initially set up \system{}, the core behavioral evaluation steps can all be done in the frontend UI by nontechnical users.

The primary tasks in behavioral evaluation are creating subsets of data and calculating relevant metrics.
The Exploration page is the initial interface for \system{} and allows users to explore, filter, and create slices of data.
It is divided into two sections, the instance view and the metadata panel. 

The instance view (\Cref{fig:exploration}, C) is a grid display of data instances, ground truth labels, and model outputs.
Users can select which model output they wish to see, which metric is calculated, and which transformation is applied to the data using the drop-down menus at the top of the UI.
A key feature of the instance view is that it is a modular Python package that supports any model and data type \textbf{(D1)}.
Each view is a separate Python package that implements a JavaScript function to render a subset of data.
While views are JavaScript functions, they are packaged as Python libraries so users can install the views they need the same way they install the \system{} package.
There are currently 6 views implemented (\Cref{fig:views}), and additional views can be created using a cookiecutter template.

The metadata panel (\Cref{fig:exploration}, A \& B) provides summary visualizations of the metadata columns and previews of user-generated data slices.
Each metadata column is shown as a row in the metadata panel, displayed with a different widget depending on what type of metadata it is.
\system{} supports 5 main metadata types: continuous, nominal, boolean, datetime, and string.
Each metadata widget is interactive and can be filtered to reactively update the instance view and other metadata widgets.
When a metadata column is filtered, the filter is shown above the instance view and the selected metric is calculated for the current subset.

When a user finds an interesting or significant subset of data, they can save the current filters as a formal slice. 
Slices can also be created in the slicing panel, which allows users to visually define and join filter predicates on metadata columns.
These slices are displayed at the top of the metadata panel with their size and the selected metric, providing a quick look at the performance for each slice.
Users can also create folders to organize their slices.

\begin{myquote}
    Emma runs \system{} to analyze her transcription model in the Exploration UI. 
    First, she filters the amplitude metadata widget and finds that the model is significantly worse at transcribing quiet audio. 
    To track this subset, she creates a slice and puts it in the \textit{audio properties} folder (\Cref{fig:exploration}, A).
    She then selects the white noise transformation and sees that the error rate increases significantly.
    She notes that they may want to augment their training data with noisy instances.
\end{myquote}

\subsection{Analysis UI: Track and Test Slices Across Models}

% OUTLINE
% P1
    % Key component is to track performance across model iterations
    % Analysis UI enables comparison and evaluation of model performance across slices and models
% P2
    % Detect regressions and high variance slices
    % Can suggest potential dataset shift or need for attention
% P3
    % Can also create "tests" for expected performance.
    % Specific reports for different types of slices.
% P4
    % Export as pngs for sharing more broadly
    
% second attempt

\begin{figure*}[t]
  \centering
  \includegraphics[width=\linewidth]{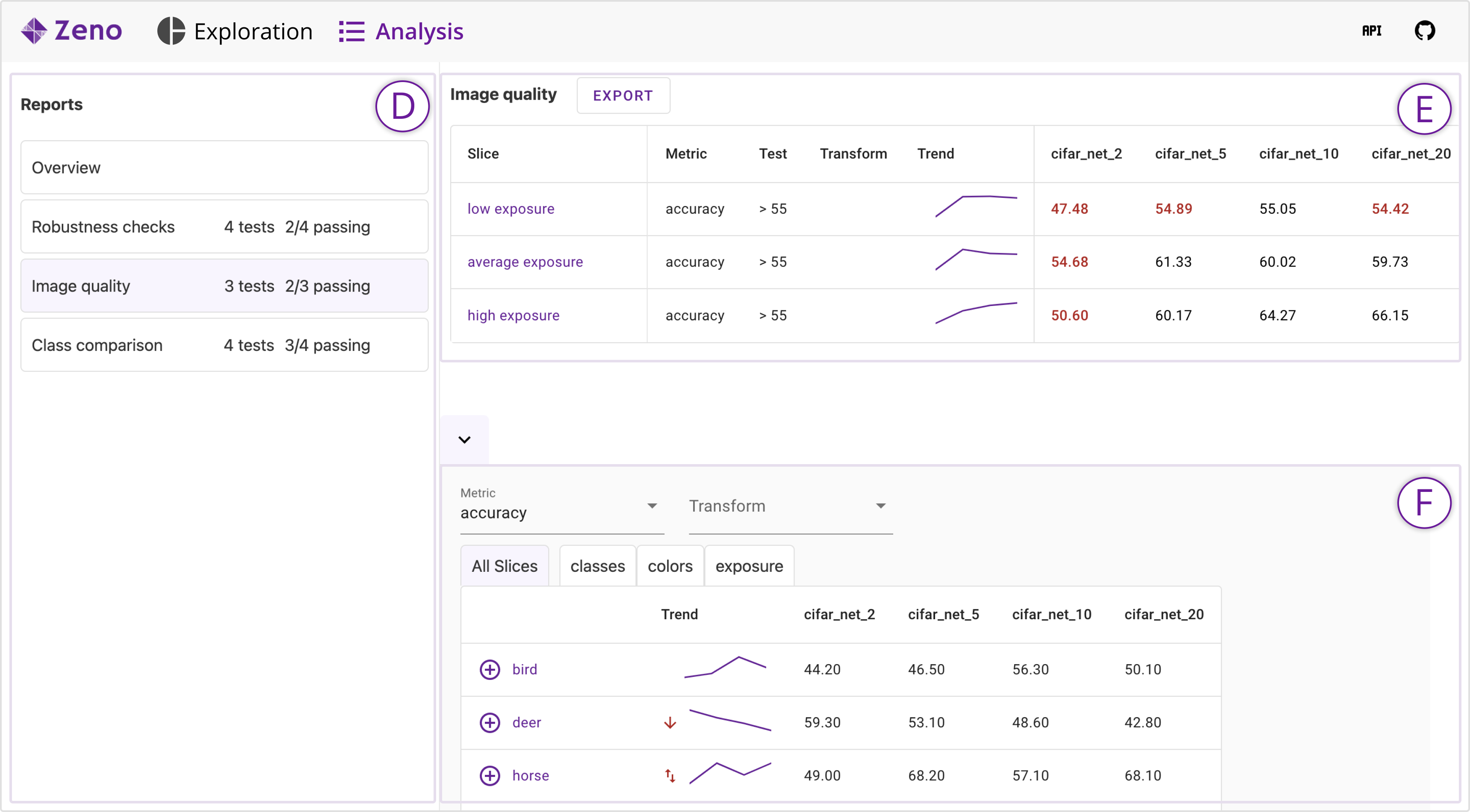}
  \caption{
    The Analysis UI helps users visualize trends of model performance across slices, and allows them to create \textit{behavioral unit tests} of expected slice metrics.
    In the figure, \system{} is shown for the CIFAR-10 image classification task comparing models trained for different epochs.
    The Slice Drawer (F) shows the performance of slices across models, including a sparkline with the metric trend over time.
    Users can create new reports in the Report Panel (D) and add slices from the Slice Drawer.
    Lastly, in the Report View (E), users can create \textit{behavioral unit tests} of expected model performance.
  }
  \Description{Image of the Zeno Analysis tab in the UI. The image shows performance on created subsets of data. It visually shows in red where the model is failing a behavioral test specified by the user.}
  \label{fig:analysis}
\end{figure*}

Once users have created the slices they wish to track using the Exploration UI, they are faced with the challenge of comparing models and slices.
The Analysis UI (\Cref{fig:analysis}) provides visualizations, reporting tools, and testing features to help users better understand and compare the performance of multiple models \textbf{(D3)}.

At the bottom of the Analysis page (\Cref{fig:analysis}, F) is a table showing the slices created in the Exploration page. 
To help users navigate the slices, folders are shown as tabs above the table and can be used to filter which slices are shown.
Users can also select which metric and transform is applied to each slice, and the resulting metric is shown as a column for each model.
To make it easier to detect trends in slice performance over time, \system{} shows a sparkline of the selected metric across models for each slice \textbf{(D3)}.

A common phenomenon for models deployed in the real world is domain shift, where the real-world data distribution changes over time and model performance degrades \cite{Moreno-Torres2012}.
To alert users of potential regressions in model performance, \system{} detects slices with performance that decreases between models.
For each slice, \system{} fits a simple linear regression of the selected metric across models, and users are alerted of slices with significant negative slope by a downward arrow next to the sparkline \textbf{(D3)}.
\system{} also highlights slices with high variance, indicating potential unexpected behavior, with a red up-and-down arrow next to the sparkline. 

Since domain shift and model updates can lead to unexpected changes in model performance, users may want to set tests for expected slice metrics.
We term these \textit{behavioral unit tests}, functions that determine whether a metric for a slice is in an expected range, such as $accuracy > 70\%$.
To create tests, users first create a new report (\Cref{fig:analysis}, D), a collection of slices, and add to it the slices they wish to test.
They can then set an expectation for a certain metric on each slice using boolean predicates on the metric value.
Models for which the test fails are highlighted in red in the report table, with the overall number of tests that failed for the most recent model shown next to each report in the report panel.
Reports can be exported as PDFs to be shared externally from Zeno \textbf{(D4)}.

\begin{myquote}
    Emma uses the insights from the Exploration UI to train a few new models with new and augmented data.
    In the Analysis UI she sees that her new models are performing better for noisy input audio, but there is a decreasing trend for instances with lower amplitude.
    To ensure that this trend does not continue, she creates a new report and adds slices for different levels of amplitude.
    She then creates behavioral unit tests expecting each slice to have an accuracy of over 65\%.
\end{myquote}

% \system{} can be hosted at a central URL for multiple stakeholders in a bigger team to access.

% first attempt
% \efedit{
% To concretely evaluate models, slices are used to create table and matrix reports. More formal png reports can be generated from the interactive page to share with diverse stakeholders and provide insights on model performance.

% From the \textit{Slices} view, slices can be found within folders or in the the "All slices" tab. An overview of each slice's performance is displayed as a sparkline and regressions or variances are highlighted through warning symbols.

% Since there can be a lot of slices and models, a table report allows users to quickly determine if the model is up to expectation across slices. To set expectations for specific slices, users add a slice to a report and adjust the metric, test, and transform. Then, from the overview of the table, under performing models are accentuated in red. 

% To analyze the relationship between sets of data, a matrix report allows users to quickly cross compare slices. Users can add slices to the X or Y axis of a matrix report and set the metric. The metric performance is displayed through the opacity of the purple indicator and the number of instances within the combination of slices is represented through the height of the purple indicator, drawing attention to any low performing or negligible combinations.
% }

\section{Case Studies}

We collaborated with four ML practitioners to set up \system{} on models they developed or audited in their work.
The goal of these case studies was to answer our second research question, whether \system{} can help practitioners working on diverse ML tasks effectively evaluate their models and discover important behaviors.
We chose these case studies as they represented a wide range of tasks (binary classification, multi-class classification, image generation) and data types (text, images, audio), testing how well \system{} generalizes.

Before each study, we met with the case study participant to understand the types of ML systems they use and decide which model(s) they wished to evaluate using \system{}.
We then worked with them asynchronously to set up an instance of \system{}, with their model, which they could access on their computer.
Finally, we conducted a one-hour study with an interview and think-aloud session (two in-person, two virtual).
During the study's first 15-30 minutes, we asked participants about their existing approaches to model evaluation and the challenges they face.
For the remainder of the study, participants shared their screen and used \system{} to evaluate the ML model, describing their thought process and findings while mentioning limitations and desired features.
Our Institutional Review Board (IRB) approved this as a separate study from the formative interviews.
In each of the following sections, we introduce the problem, describe the participant's existing evaluation approach, and detail their findings from using \system{}.

\subsection{Case 1: UI Classification}\label{sec:ui}

For the first case study, we worked with a researcher developing a model to classify smartphone screenshots using a CNN-based deep learning model, which they were evaluating on 10,000 images.
The model aims to make UIs more accessible to people with visual impairments by informing them of the type of interface they are looking at.
The participant was looking to expand their system to screenshots from other devices, e.g., tablets, and wanted to understand their model's current performance and generalizability.
Uniquely for this case study, the participant ran \system{} on a cloud server that hosted their data and models and they accessed the \system{} UI remotely on their laptop.

\subsubsection{Existing evaluation approach.}
The first participant primarily uses computational notebooks for both \textit{qualitative} and \textit{quantitative} evaluation of their models.
For \textit{qualitative} analyses, they select \feedback{some test cases that I hypothesized are hard and easy for the model}, instances for which they check the model's output to understand how it is behaving.
For example, for this model they check a specific screenshot of a login screen with a list structure that they expect the model to misclassify as a list view. 
For every new domain in which they train a model, the participant spends significant time creating dedicated Python notebooks to display data instances and model outputs for this type of qualitative analysis.

The participant also uses \textit{quantitative} metrics for evaluation, especially for more complex domains such as object detection where they use a combination of metrics such as mean Average Precision (mAP) at different scales.
As with the qualitative analyses, the participant authors specific Python notebooks to calculate these metrics.
They also make an effort to write evaluation code that is distinct from the training code to ensure that they avoid any bugs such as data leakage in the training process.
% \efcomment{is there an example of an interface like this, could we see the ones that he hypothesized the model would fail on?} 
% \apcomment{Would it make senses to be more specific about their painful process without Zeno?}
% \jason{Yeah, this kind of before / after would be really useful. How much effort, how long, how difficult, etc.}

% \accomment{something about jason's comment}.
% \apcomment{Can you describe HOW they used Zero?}
% They were also able to discover new hypotheses by browsing the instance view and found it especially useful to filter and look through incorrectly classified instances.
% \jason{So if I understand correctly, previously, they were just eyeballing things rather than doing it systematically. This also means that they hadn't set up regression tests, to make sure functionality doesn't break as new models are trained. Is this right? This might be a point worth emphasizing more, that setting up regression tests properly and tracking over time is a real pain, and there aren't tools for that}

\subsubsection{Findings with \system{}.}
The participant found \system{}'s interactive instance view and metadata distributions extremely useful for discovering new failures, systematically validating qualitative analyses, and sharing results with others.
Just from the initial Exploration UI, the participant found the ability to quickly browse dozens of instances much more valuable than the static notebooks they used previously.
Within a few seconds, they found new model failures they noted to validate later and add as new qualitative test examples. 
The participant wished to filter the instance view to only see failures or have the system suggest slices to make it easier to quickly find model errors.

\begin{figure*}[t]
  \centering
  \includegraphics[width=\linewidth]{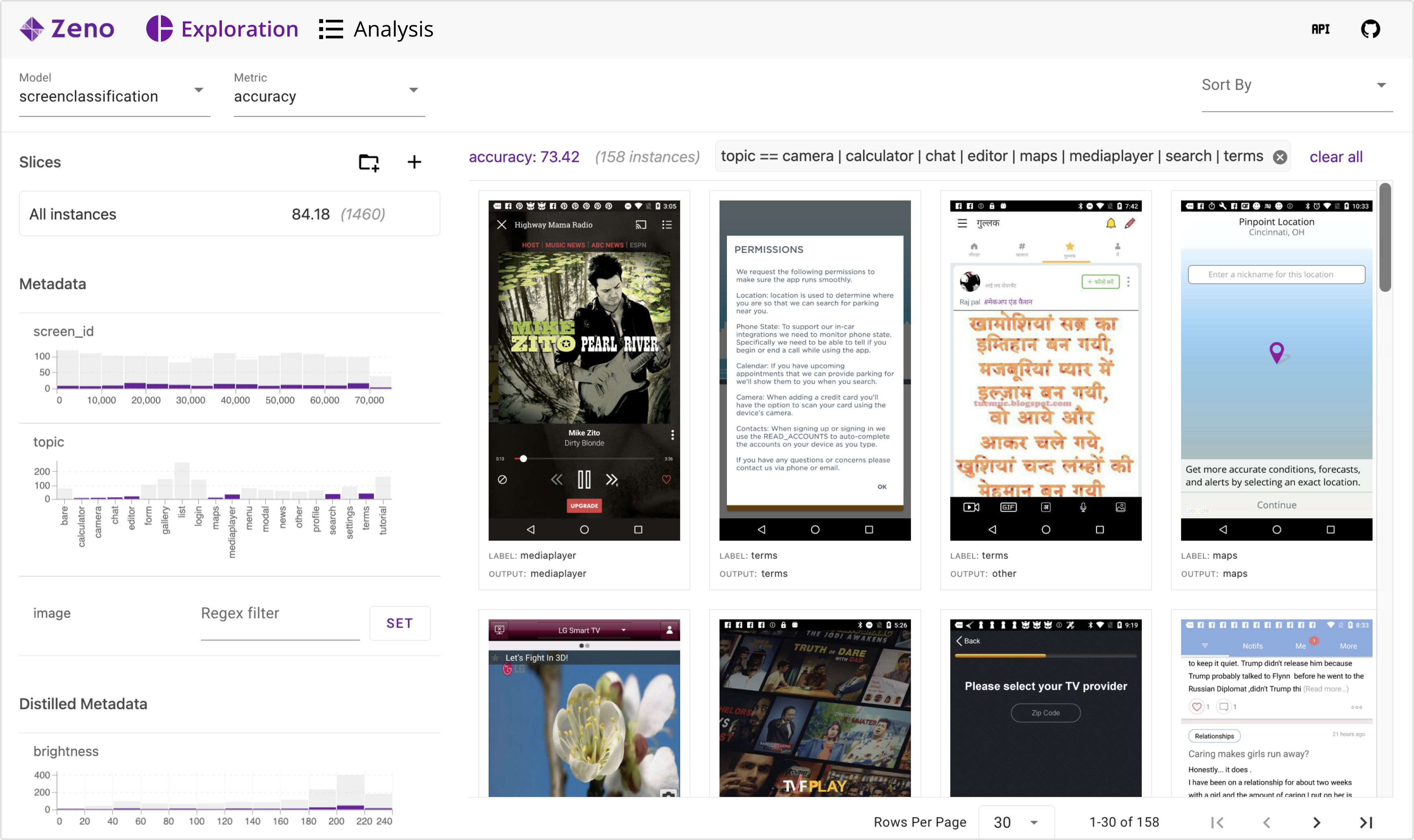}
  \caption{
    A screenshot of the Exploration UI from the UI classification case study (\Cref{sec:ui}).
    The participant selected underrepresented ground-truth classes and confirmed that the model performance is significantly worse for them.
  }
  \Description{An image of the Zeno Exploration UI showing the case study for UI classification. The image shows metadata on the sidebar and data consisting of UI screenshots in the main view.}
  \label{fig:case}
\end{figure*}

With the metadata distributions in the Exploration UI the participant was also able to validate some of their existing qualitative hypotheses more systematically.
For example, they confirmed their hypothesis that the model would perform worse for underrepresented classes in the dataset by filtering for the most underrepresented classes using the class histogram (see \Cref{fig:case}). 
They found the ability to save such slices of data to share with others to be a powerful feature and wished to
\feedback[]{take a very well known dataset such as ImageNet, find slices that are questionable and share them} to help others test their own model for such issues.

Lastly, the participant found that the code for the \system{} API was similar to what they used in notebooks and that they \feedback[]{could totally get used to the \system{} API}.
While they were able to copy and paste their existing code into \system{}, they wished for a more streamlined setup process, for example, with automatically generated \system{} configuration files for common data types and ML libraries.

% Although the participant found the \system{} interface intuitive and similar to gallery views in consumer applications, they had some suggestions for further improvement.
% Their main usability concern was that they did not originally know that they could click and drag the metadata distributions to filter their data.

% \apcomment{Feels a bit shallow and vague as written?  You have a screenshot of their use case but you don't integrate it into the description of what they did.  This would make it feel more tangible.}
% \jason{Another thing that could help, try to convey size of data. Are we talking about 100s, 1000s, 10000s? Same comment for other case studies.}

\subsection{Case 2: Breast Cancer Detection}\label{sec:cancer}

In the second case study, we worked with a researcher who was auditing a breast cancer classification model on a dataset of 6,635 images.
The model, also a CNN-based deep learning model, divides mammogram images into small patches and detects whether there is a lesion present in each patch.
The model was trained on a dataset provided by a collaboration with clinical researchers at an academic hospital system in the United States.
Although the model had a reasonably high accuracy of 80\%, the developers had difficulty understanding the failure modes of the model, especially since the dataset was de-identified and had minimal metadata.
The participant in our case study wanted to discover meaningful dimensions across which the model failed in order to guide model updates.
% the University of Pittsburgh Medical Center (UPMC), in collaboration with clinical researchers there \apcomment{think if we should anonymize this}.
% The participant was verifying the performance of the model and checking how it performed in potentially important edge cases.
% \apcomment{I think we need to motivate this more about why Zeno is needed.  The bigger story is that the overall accuracy of this model was around 80\% but researchers couldn't figure out slices of data where the model was consistently failing.  This meant that without understanding this, they didn't have direct evidence on how to improve the model as 80\% is good but maybe not good enough for clinical impact.   There was very little metadata associated with these images, given the images were de-identified, which is why they needed to derive features.   Perhaps you can contrast this with the previous study which had lots of meaningful metadata.}

\subsubsection{Existing evaluation approach} 
Unlike the first case study participant, the participant in the second study had only used quantitative aggregate metrics when evaluating models.
They \feedback{had not used any platform or framework to understand how a model performed on specific features of the metadata}, and fully relied on aggregate metrics as a measure for model quality.
This involved creating Python scripts to load a model and data and calculate metrics such as AUC and F1 score.
Attempting to improve the breast cancer classification model led to their first foray into behavioral evaluation.

\subsubsection{Findings with \system{}}
The participant found that the combination of the extensible \python{@distill} functions and metadata distributions was essential for finding slices with significant areas of error. 
Since the participant was not a domain expert, they consulted with medical imaging researchers that recommended a Python library, pyradiomics \cite{van_griethuysen_computational_2017}, to extract physiologically relevant characteristics from medical images.
% The participant started their evaluation by looking at some simple image-based features, such as brightness, but did not find any patterns that impacted performance.
The participant implemented dozens of \python{@distill} functions using pyradiomics functions that encoded important regional information, such as grey-level values, that was not captured by their original features.
They also wrote a couple more \python{@distill} functions to encode the position of each image patch, a hypothesis they had from looking at model failures in the instance view.
The participant only had to add a couple of lines of Python to use all of these functions in \system{}.
% \jason{Also, was this easy to do? Showing that Zeno (again) streamlines well and integrates easily is a plus. It seems that this can be broken into several steps, e.g. installation (which you took care of), importing data into Zeno, importing a model into Zeno, writing initial functions, doing initial analysis, creating slices, iteration. So saying that it not only streamlined, but that people got the expected Zeno workflow really quickly is a plus. For Discussion, if everyone got into the expected Zeno workflow quickly, then that feels like a good win.}

Since the dataset had minimal existing metadata, interactively filtering the \python{@distill}ed distributions was the primary way the participant found patterns of failure.
By interactively cross-filtering the \python{@distill}ed metadata histograms, they found that the model performed significantly worse for images with higher tissue density, a phenomenon that also occurs with human radiologists \cite{kolb_comparison_2002}.
They also found that the model was trained on many background patches of image that did not include part of the breast, which also impacted the aggregate metrics.
The participant noted that they may want to clean the data and upsample instances relevant to the classification task.
Due to the quantity and complexity of these analyses, the participant wished for more expressive slice comparisons, such as comparing multiple slices at a time in the Exploration UI.
Otherwise, using \system{} the participant found significant failures which they had not been able to find using Python scripts.
% \apcomment{Can we talk about how the interactivity of filters led to such discoveries?  Would have been hard to arise at this combination otherwise.}

% As the model was trained on image patches, the participant was curious if the position of the patch had any impact on the model's accuracy.

% Currently, the model was trained by selecting random patches from breasts without lesions.  
% However, training on more patches from the central region of the breast would ensure the model would have more dense and texturized images. 
% The participant believes that re-training the model on these patches could help the model generalize more effectively.

% Although the participant had not conducted behavioral evaluation before, they found \system{} to be intuitive and were able to quickly discover ways in which the model could be improved.
% One limitation they faced was creating more complex slices consisting of nested predicates such as 
% $(A  \&\&  B) || (C || D)$, and we worked with the participant to expand the slice creation UI to support nested queries.

\subsection{Case 3: Voice Commands}\label{sec:dov}
The third case study was with a participant who was developing a decision-tree model to detect the direction in which a person is speaking using an array of microphones, which they were evaluating on 11,520 recordings.
The goal of the model is to predict to which microphone, often a smart speaker, a person is talking in order to respond from the right speaker.
The participant had collected data from diverse setups to understand the performance of their model in the different scenarios.

\subsubsection{Existing evaluation approach}
Most of the models the participant works on are sensor-based systems highly impacted by the physical nature of the data signals, for example, echoes and noise in sound data.
Thus, in addition to calculating classic aggregate metrics, the participant generates and tests inputs with diverse physical properties.
For example, in the model described above, the participant collected audio from speakers next to a wall and in the middle of the room to since they thought the rebounding sound from the wall might confuse the model.

To evaluate such scenarios, the participant collects data in dozens of configurations, and so often has extensive metadata for behavioral analysis.
Like the other participants, they use computational notebooks to manually split the data across different metadata features and print out multiple metrics.
Due to their high quantity of metadata, the participant only looks at simple slices of data, and does not often explore intersectional slices of multiple features.

% The participant also emphasized the importance of a fast iteration loop of testing hypotheses, collecting data, and retraining their model.

\subsubsection{Findings with \system{}}
Using \system{}, the participant was both able to validate all of their hypotheses significantly faster and discovered potential causes for systematic model failures.
For example, they confirmed a finding from previous analyses where a \feedback[]{model worked very well at 1, 2, and 3 meters, but there was a sharp dropoff at 5 meters} by simply looking at the metadata distributions. 
They also used the spectrogram visualization of instances in each slice to look for potential reasons for the steep dropoff in performance, for example, signals with lower amplitude. 
Additionally, they found the cross-filtering between metadata histograms to be useful to find potential interactions between physical features, such as audio both at a distance and a speaker against a wall.
Cross-filtering combined with expressive instance visualizations of the audio data was essential for both confirming their hypothesis and ideating potential causes for model failures.

Much of the participant's work is focused on collecting new data, so they suggested data-related improvements for \system{}.
Since the participant often tests their model with their own inputs, they wished for a direct way to add new instances to \system{}.
They also mentioned having more interactive transformations, for example, having a slider to gradually apply a transformation such as reducing the amplitude of an audio file.

\subsection{Case 4: Text-to-Image Generation}\label{sec:diffusion}

For our last case study, we worked with a non-technical researcher who explores biases in deployed ML systems, in this case, the text-to-image generation model Stable Diffusion \cite{rombach_high-resolution_2022}.
To audit this model they used \system{} with the DiffusionDB Dataset \cite{wang_diffusiondb_2022}, which consists of 2 million prompt-image pairs generated using the Stable Diffusion model.
The participant wanted to explore potential systematic biases in the images generated by Stable Diffusion.

\subsubsection{Existing evaluation approach}
The participant's work is primarily focused on auditing public-facing algorithmic systems such as search engine results and social media ads.
They exclusively conduct manual, ad-hoc audits, testing a range of specific inputs such as search queries and individually checking the model's outputs.
The inputs they test are often guided by existing knowledge of model biases, for example, the participant has \feedback[]{used some lingustic discrimination knowledge [...] such as knowing that certain words tend to be gendered} to test inputs with likely biased results.

The participant also works with end users of algorithmic systems to understand how they audit models and what biases they are able to find. 
They found that \feedback{people often found issues in searches that none of the researchers, including me, had even thought of}.
Having diverse users test models is essential for finding issues, and the participant works with end-users to surface new limitations.

% ``Ad for men on a videogame, could be biased if systematic''. 

\subsubsection{Findings with \system{}}
When auditing the DiffusionDB dataset with \system{}, the participant took a similar approach to their previous audits but was able to come up with more systematic and validated conclusions of model biases.
Their primary interaction with \system{} was using the string search metadata cell to look for certain prompt inputs.
Similar to how they approached debugging search engines, they used prior knowledge of likely biased prompts but were able to see dozens of examples instead of one prompt at a time.
For example, when searching for prompts with the ``scientist'' in them, every generated image was male, encoding a typical gender bias.
By seeing dozens of prompts the participant was able to gather more evidence that the model produced this pattern systematically and was not due to a one-off prompt.

The DiffusionDB dataset also includes a measure of toxicity, or ``NSFW'' level, for both the input prompts and generated images.
These numbers were represented as histogram distributions in \system{}, and the participant found it invaluable to filter by and find potential biases.
One interesting experiment the participant tried was to see if the average distribution of the NSFW tag would go up for certain terms.
For example, they saw small increases in the distribution when searching for certain gendered terms, including the word ``girl'', which reflected that the images generated of women were more sexualized than those of men.
They could only see this dataset-level pattern using the combination of \system{}'s metadata distribution and instance view.

Lastly, the participant reflected on how usable \system{} would be for everyday users of algorithmic systems. 
They mentioned that technical terms such as ``metadata'' may be too niche for everyday users and could be renamed.
Otherwise, they found the system intuitive and usable if set up for use by diverse end users.

\section{Discussion}
Our case studies showed that \system{}'s complementary API and UI empowered practitioners to find significant model issues across datasets and tasks.
More generally, we found that a framework for behavioral evaluation can be effective across diverse data and model types \textbf{(D1)}.
This generalizability can be seen by comparing two of the case studies, the malignant tumor detection (\Cref{sec:cancer}) and audio classification (\Cref{sec:dov}) cases.
The two cases differed significantly in their data type (image vs. audio), task (binary vs. multi-class classification), model (CNN vs. decision tree), and end goal (model development vs. auditing).
Despite these differences, both participants could effectively discover and encode model behaviors they wished to test and found limitations ranging from robustness to domain shift \textbf{(D2)}.

\system{}'s different affordances made the behavioral evaluation process easier, quicker, and more effective, depending on the user's goals and the challenges of each particular task.
% The participants found the \system{} UI to be intuitive \textbf{(D4)} and powerful \textbf{(D2)}.
For example, in Case 2, the participant found the extensible API essential for creating metadata to analyze their model across \textbf{(D2)}, while in case 3, the participant found the interactive visualizations more useful given the extensive metadata already present in their dataset.
\system{} also supports users' particular strengths and skillsets - without using the API, our non-technical case study participant (Case 4) was still able to find significant model biases by using their domain knowledge to interact with the UI \textbf{(D4)}.
% The combination of aggregate metrics and individual model predictions also turned out to be powerful, leading to analyses that were more fine-grained than overall metrics and more evidence-backed than looking at individual predictions.

Participants in the case studies found that \system{} was easily integrated into their workflows, requiring minimal effort to adapt their code to work with the \system{} API \textbf{(D1)}.
For example, the participant in case study 1 only modified a few lines of their inference code to work with \system{}, and the participant in the second case study was able to use a radiomics library in \system{} with minimal setup.
The participants also suggested ways in which \system{} could be made even easier to use, such as automatically generating \system{} API functions and configuration files for common ML libraries.

% \efcomment{maybe talk about how zeno allows users to understand their data like they can with most currently systems, but beyond that they can also contextualize their model's performance through instances and semantic slices that anyone can understand}
% Overall, \system{} empowered users to conduct more systematic and principled evaluations.
% Participants generally described two distinct evaluation processes, quantitatively checking the overall performance of a model using aggregate metrics and qualitatively sanity checking a model's behaviors by spot-checking specific instances.
% \system{}'s slice-based evaluation bridges the two types of evaluation, with insights that are more fine-grained than overall metrics and more evidence-backed than looking at individual predictions.

While we validated that most of the design goals were met by \system{}, our case studies did not thoroughly explore how \system{} could be used over longer periods \textbf{(D3)}.
All four participants worked with early-stage models and only used \system{} for a limited time.
Longer-term, in-situ studies would provide more nuanced feedback for the utility of \system{}'s model comparison features.
A benefit of \system{}'s ease of use, both with the API and UI, is that users can immediately start using \system{}'s model tracking and comparison features as models move from research to deployment.

\section{Limitations and Future Work}

% OUTLINE
% P1 - Discovery
    % hard problem
    % algorithmic methods
    % crowdsourced method
    % plug into zeno
% P2 - visualizations
    % not very good in zeno
    % instance views
    % performance views
% P3 - better for non-programmers
    % default packages/setups without code
% P4 - updating models
    % slice-based learning
    % GDRO

\system{} provides a general and extensible framework for the behavioral evaluation of ML, but leaves significant room to better address the challenges in the evaluation process.

\vspace{1em}
\noindent\textit{Slice discovery.}
A central challenge for behavioral evaluation is knowing \textit{which} behaviors are important to end users and encoded by a model.
To directly encourage the reuse of model functions to scaffold discovery, we are currently designing \textit{ZenoHub}, a collaborative repository where people can share their \system{} functions and find relevant analysis components more easily.
Including slice discovery methods directly in \system{} could also help users find important behaviors.
\system{} provides the common medium of representing metadata and slices that practitioners can use to interact with and use the results of these discovery methods.

\vspace{1em}
\noindent\textit{Improved visualizations.}
Defining and testing metrics on data slices is the core of \system{}, but it only provides a few simple visualizations of data and slices in a grid and table view.
There are many more powerful visualization types that could improve the usability of \system{}.
Instance views that encode semantic similarity, such as DendroMap \cite{bertucci_dendromap_2022}, Facets \cite{Pushkarna2017}, or AnchorViz \cite{chen_anchorviz_2018}, could improve users' ability to find patterns and new behaviors in their data.
\system{} can also adapt existing visualizations of ML performance, such as ML Cube \cite{Kahng2016}, Neo \cite{gortler_neo_2022}, or ConfusionFlow \cite{Hinterreiter2020}, to better visualize model behaviors.
For example, grid views showing the intersections of slices could highlight important subsets of data.

\vspace{1em}
\noindent\textit{Scaling.}
\system{} has a few optimizations for scaling to large datasets, including parallel computation and caching, but machine learning datasets are continuously growing and additional optimizations could speed up processing considerably.
A potential update would be to support processing in distributed computing clusters using a library such as Ray \cite{moritz_ray_2018}.
Another bottleneck is the cross-filtering of dozens of histograms on tables with millions of rows.
\system{} could implement an optimization strategy like Falcon \cite{moritz_falcon_2019} to support live cross-filtering on large datasets.

\vspace{1em}
\noindent\textit{Model improvement.}
\system{} is focused exclusively on \textit{evaluation} and does not include methods to update models and fix discovered failures.
Future work can explore how to directly use the insights from \system{} to improve model performance.
For example, there are promising results in using data slices to improve model performance, such as slice-based learning \cite{Chen2019a} and group distributionally robust optimization (GDRO) \cite{sagawa_distributionally_2020, liu_just_2021}.

\vspace{1em}
\noindent\textit{Further evaluation.}
The case studies evaluated \system{} on real-world ML systems, but further evaluations could better elucidate the affordances and limitations of \system{}.
Future evaluations could explore how usable \system{} is for 
additional non-technical users and how well it works for continually updated deployed systems.

\section{Conclusion}

% outline
    % Behavioral evaluation important
    % Zeno good for this
    % As a foundation for this, can integrate more powerful tools
    
Behavioral evaluation of machine learning is essential to detect and fix model behaviors such as biases and safety issues.
In this work, we explored the challenges of ML evaluation and designed a general-purpose tool for evaluating models across behaviors.

To identify specific challenges for ML evaluation, we conducted formative interviews with 18 ML practitioners.
From the interview results we derived four main design goals for an evaluation system, including supporting comparison over time and no-code analysis.
We used these goals to design and implement \system{}, a general-purpose framework for defining and tracking diverse model behaviors across different ML tasks, models, and data types.
\system{} combines a Python decorator API for defining core building blocks with an interactive UI for creating slices and reports.

We showed how \system{} can be applied to diverse domains through four case studies with practitioners evaluating real-world models.
Participants in the case studies confirmed existing findings, hypothesized new failures, and validated and discovered behaviors using \system{}.
As a general framework for behavioral evaluation, \system{} can incorporate future features, such as error discovery methods and visualizations, to support the growing complexity of models and encourage the deployment of responsible ML systems.

\begin{acks}
We would like to thank Fred Hohman, Alex Baüerle, Will Epperson, and Dominik Moritz for their feedback.
This material is based upon work supported by a Mozilla Technology Fund grant, a Cisco Research Grant, an Amazon Research Award, a National Science Foundation grant under No. IIS-2040942, and the National Science Foundation Graduate Research Fellowship Program under grant No. DGE-1745016. Any opinions, findings, and conclusions or recommendations expressed in this material are those of the authors and do not necessarily reflect the views of the grantors.
\end{acks}

\bibliographystyle{ACM-Reference-Format}
\bibliography{alex-refs}


\end{document}